\documentclass{aa}  

\usepackage{graphicx}
\usepackage{txfonts}
\usepackage[]{hyperref}
\usepackage{natbib}
\usepackage{color}
\usepackage{xcolor}
\bibpunct{(}{)}{;}{a}{}{,} 
\begin{document}

   \title{Prospects for detection of the pair-echo emission from TeV gamma-ray bursts}


   \author{D.Miceli
          \inst{1}
          \and
          P. Da Vela
          \inst{2}
          \and
          E. Prandini
          \inst{1}
          }

   \institute{INFN Sezione di Padova and Università degli Studi di Padova, Via Marzolo 8, 35131 Padova, Italy\\
              \email{davide.miceli@pd.infn.it}
         \and
            INAF - Osservatorio di Astrofisica e Scienza dello spazio di Bologna, Via Piero Gobetti 93/3, 40129 Bologna, Italy \\
            \email{paolo.davela@inaf.it}
             }

   \date{}

 
  \abstract
{The intergalactic magnetic field (IGMF) present in the voids of large-scale structures is considered to be the weakest magnetic field in the Universe. Gamma-ray observations of blazars in the GeV-TeV domain have led to lower limits on the IGMF strength based on the search for delayed or extended emission. Nevertheless, these results have been obtained with strong assumptions placed on the unknown source properties. The recent discovery of TeV radiation from gamma-ray bursts (GRBs) has paved the way for IGMF studies with these bright transients. Among the current TeV-detected GRBs, GRB\,190114C, located at a redshift of $z = 0.42$, is the best sampled. Therefore, it can be considered to be representative of the properties of GRBs in the VHE domain. In addition, { GRB\,221009A ($z = 0.151$) is the brightest event ever detected.}
We present a phenomenological model based on the intrinsic properties of GRB\,190114C {and GRB\,221009A} to predict the delayed emission component (pair-echo) in the GeV-TeV band. We investigate the detectability of this component from low-redshift ($z \leq 1$) GRBs {for three values of IGMF strength ($10^{-19}$ G, $10^{-18}$ G and $10^{-17}$ G)}, different observational times ($3$ hrs, $6$ hrs, and $9$ hrs) and source intrinsic properties. 
We find that {for} current and future generation $\gamma$-ray instruments, {extending the observation for at least 3 hours after the GRB detection is a viable strategy for probing the IGMF. We also confirm that GeV-TeV observations of GRBs can probe IGMF strengths on the order of $10^{-17} -10^{-19}$ G, representing a competitive alternative to the current studies performed with active galactic nuclei (AGNs)}.} 

   \maketitle
%

\section{Introduction}
\label{sec:intro}
The presence of magnetic fields in structured regions of the Universe, such as galaxies or galaxy clusters, has been clearly verified by observations \citep{beck96,moss96,kim1990,carilli02}.
The origin of these fields is debated, with an astrophysical origin posited against a cosmological (or primordial) one. Evidence of the presence of weak magnetic fields in the intergalactic medium could help probe the latter theory.
In this scenario, the galactic and cluster micro-Gauss magnetic fields are generated through the dynamo amplification of primordial magnetic seeds \citep{kronberg94,widrow02,widrow12}. This implies that the voids between galaxies and galaxy clusters contain an intergalactic magnetic field (IGMF) generated in the early Universe. 
The identification of IGMF properties can provide direct information on magnetogenesis and amplification processes. This can have an impact on several fields of astrophysics, particle physics, and cosmology \citep[see, e.g.,][]{durrer13,subramanian16}.

Promising approaches developed to probe IGMF properties (strength, $B$, and coherence length, $\lambda_B$) exploit gamma-ray data in the GeV-TeV range \citep{neronov09}.
Very high energy (VHE, E $> 100$ GeV) photons coming from an extragalactic source interact with extragalactic background light (EBL), generating electron-positron pairs. These particles can then up-scatter the CMB photons via inverse Compton generating a secondary radiation component up to tens of GeV. The energy of these reprocessed secondary photons $E_{rep}$ is related to that of the source emitted photons $E_{\gamma_0}$ following the relation $E_{rep} \simeq 0.32 (E_{\gamma_0}/ 20\text{ TeV})^2$ TeV \cite{neronov09}.
 In the presence of an IGMF, the electron-positron pairs are deflected. As a result, the detected secondary radiation might be extended (pair-halo emission) and/or time-delayed (pair-echo emission) with respect to the primary emission coming from the extragalactic TeV source. 
 Various methods have been devised to assess the spectral \citep{neronov10}, angular \citep{dolag09,ando10,chen15}, and temporal \citep{plaga95,murase08} characteristics of this component. 

Almost 100 extragalactic emitters have been detected as point-like sources at TeV \footnote{\url{http://tevcat2.uchicago.edu/}}.
Several studies have been performed exploiting persistent TeV emission observed from active galactic nuclei (AGNs) \citep{neronov10,taylor11,dermer11,tavecchio11}. Lower bounds on the IGMF strength have been derived in the range of $\sim$ $10^{-15}-10^{-17}$ G from deep GeV-TeV observations under different assumptions on the coherence length or intrinsic spectrum shape and flux variability of the primary source \citep{ackermann18,aleksic10,veritas17,hess14}. {Most} of these results suffer from the strong assumption on the knowledge of the intrinsic (hence, not reprocessed) source properties and its variation in time and the source duty cycle. {A recent study \citep{magic22} put a lower bound of the order of $\sim$ $10^{-17}$ G for large correlation lengths from long-term observations of 1ES\,0229+200, relaxing these assumptions and taking into account also the variability of the emission in the TeV energy band}.

The recent discovery of radiation coming from gamma-ray bursts (GRBs) in the VHE $\gamma$-ray band \citep{magic19_14C,hess18} opens up the possibility of also testing the properties of IGMF with transient $\gamma$-ray sources, as proposed in \citet{razzaque04,ichiki2008,takahashi08,murase09}. The IGMF strengths explored with the pair-echo emission from GRBs are typically weaker ($10^{-17}-10^{-21}$ G) compared to the lower bounds derived by AGNs. Nevertheless, in this case, some of the assumptions that affect the study of IGMF with AGNs can be relaxed, for instance, the source duty cycle. The possibility to reduce the number of assumptions {and to have an independent verification of the bounds derived with AGNs} justifies the exploration of the IGMF strengths {with GRBs}. Overall, GRB emission that extends even above 1 TeV \citep{hess_190829A,lhaaso22,lhaaso23} has made this opportunity very appealing. As a result, the events showing long-lasting $\gamma$-ray emission as in the case of GRB\,130427A, GRB\,190114C and GRB\,221009A have been proposed for the first IGMF studies on transient sources. 

In \citet{veres17} an extrapolation of GRB\,130427A spectrum to TeV energies shows the feasibility of \textit{Fermi}-LAT and VERITAS in constraining the IGMF in the case of similar events for magnetic strength in the $10^{-21}-10^{-17}$\,G range assuming a coherent length of 1\,Mpc. 
From the comparison between \textit{Fermi}-LAT upper limits and the scenario assumed for a GRB\,130427A-like event at TeV energies, the IGMF strength in the $10^{-20}-10^{-18}$\,G range was ruled out.

Recently, \citet{wang20} derived a lower bound on the IGMF strength of $10^{-19.5}$ G for a coherent length $\leq 1$ Mpc for GRB\,190114C. \textit{Fermi}-LAT upper limits for 15 days, 1 month, and 3 months of observational time were compared with the expected secondary emission. However, a different conclusion was reached in \citet{dzhatdoev20,paolo_paper}. In these latter studies, the \textit{Fermi}-LAT upper limits on GRB\,190114C were shown to be consistent with the expected observable cascade emission for IGMF strength, ranging from $10^{-21}$\,G up to $10^{-19}$\,G. Therefore, no constraints on the IGMF strength and/or structure could
be claimed using this dataset. The discrepancy is explained in terms of different approaches in the assumptions of the primary source model and in the secondary emission calculation techniques. In \citet{2023vovkonly} the Fermi-LAT gamma-ray emission at late time ($10^4$\,s after trigger time) is interpreted as the pair-echo emission in the case of an IGMF strength $B \leq 10^{-21}$ G.

Similar studies have also been performed for GRB\,221009A \citep{2023ApJ...955L..10H, 2023vovk,2022wang, 2024dzhatdoev,2023mirabal}, using \textit{Fermi}-LAT observations. IGMF lower limits ranging from $10^{-20}$\,G up to $10^{-16}$\,G have  been derived. Overall, the most recent studies have mostly focused on the pair-echo emission expected at energies below a few tens of GeV that can be probed by space-born instruments; in particular,  \textit{Fermi}-LAT. The possibility and capabilities of Cherenkov instruments working in the VHE $\gamma$-ray band have not been clearly explored up to now. As a result, a conclusive answer on the capability of current or future instruments in the GeV-TeV band at the current stage is far from achievable. However, these preliminary studies have shown that GRBs seem to be promising sources for IGMF studies. 

In this article we explore the possible observational strategies for the current and future generation VHE $\gamma$-ray detectors to perform IGMF studies.
Starting from the intrinsic properties of GRB\,190114C, we  build a phenomenological model to estimate the expected secondary cascade emission from GRBs and compare it with the performances of the MAGIC telescopes and the CTA-North array. {We also explore the exceptional case of GRB\,221009A, estimating the capabilities of the MAGIC and CTA-North array for this extremely bright event}. In Sect. \ref{sec:methods} the phenomenological model{s} adopted to describe the intrinsic properties of GRBs and the estimated secondary cascade are presented. In Sect. \ref{sec:results} we discuss the comparison between the secondary cascade in the GeV band and the sensitivity of Cherenkov instruments for transient observations, assuming {three different IGMF strength values}, observational times, and source intrinsic properties. Finally, in Sect. \ref{sec:conclusion} we examine the contribution of future VHE $\gamma$-ray detectors and the benefit of IGMF studies from GRBs. In this work, we assume the cosmological parameters of $H_0=70$ km s$^{-1}$ Mpc$^{-1}$, $\Omega _{\Lambda} = 0.7$, and $\Omega _{M} = 0.3$. \section{Phenomenological model for GRB emission}
\label{sec:methods}

\subsection{GRB\,190114C}\label{subsec:190114C}
The long gamma-ray burst GRB\,190114C was triggered by \textit{Swift}-BAT \citep{swift_gcn_190114C} on 14 January 2019 at $T_0=20$:$57$:$03$\,UT, and by the \textit{Fermi}-GBM \citep{fermigbm_gcn_190114C} instruments. Several other space $\gamma$-ray instruments such as AGILE, INTEGRAL/SPI-ACS, Insight/HXMT, and Konus-Wind \citep{magic_14C_modeling} detected the event. The light curve recorded by these instruments showed a multi-peak structure, typical of the prompt phase, only for about 25\,s. This suggests that the remaining activity, which is characterized by a smooth power-law decay, may be dominated by afterglow emission. Support for this interpretation was also obtained from a joint spectral and temporal analysis of the \textit{Fermi}-GBM and \textit{Fermi}-LAT data, where the afterglow onset was estimated to be at $\sim 6$ s \citep{ravasio_190114C}. In terms of energetics, the isotropic-equivalent energy is $E_{\gamma,iso} = (2.5 \pm 0.1) \times 10^{53}$\,erg in the energy range 1-10$^4$\,keV \citep{fermi_swift_190114C}, which places this GRB in the range of the $20\%$ most energetic ones, when considering the long GRBs catalog by \textit{Fermi}-GBM with known redshift \cite{fermi_gbm}.

Multi-wavelength follow-up observations were performed covering the entire electromagnetic spectrum from radio up to TeV energies. As a result, the redshift was estimated to be $z = 0.4245 \pm 0.0005$ \citep{not_190114C,gtc_190114C}.
Follow-up observation in the VHE $\gamma$-ray band was performed by the MAGIC telescopes \citep{magic19_14C} from $\sim60$\,s after $T_0$. A strong signal was recorded with significance above $50 \sigma$ in the first 20 minutes of observation {\citep{magic19_14C}}. The event was detected up to 2454\,s after the trigger time $T_0$.
 The light curve in the $0.3-1$\,TeV range corrected for EBL absorption with the EBL model \citet{dominguez11} in the $62-2454$\,s time interval is described by a power law with a temporal decay index $\beta_T = - 1.60 \pm 0.07$. A constant value of $\alpha_{int}$ $\approx -2$ is estimated for the intrinsic (i.e., EBL-corrected) spectral photon index for the entire detection period.  Nevertheless, evidence for a softening of the spectrum with time is not completely excluded.
The power law spectral fit in the $0.2-1$\,TeV energy range for the time-integrated emission $62-2454$\,s returns $\alpha_{obs} = -5.34 \pm 0.22$ and $\alpha_\textup{int}=-2.22^{+0.23}_{-0.25}$ for the observed and EBL-corrected spectrum, respectively.

\subsection{GRB\,221009A}
{
\label{subsec:221009A}
GRB\,221009A is an extremely energetic ($E_{\gamma,iso} \sim 1 \times 10^{55}$ erg), bright ($L_{\gamma,iso} \sim 9.9 \times 10^{53}$ erg/s), and nearby ($z = 0.151$) event triggered by \textit{Fermi}-GBM on 9 October 2022 at 13:16:59.99 UT \citep{221009A_gcn_gbm,221009A_fermi_gbm}. The event was subsequently detected also by the \textit{Fermi}-LAT instrument \citep{2022GCN.32637....1B,2022GCN.32658....1P}, which records photons up to $\sim 100$\,GeV and by the \textit{Swift} satellite \citep{2022GCN.32632....1D} both by BAT and XRT instruments, with  observations starting $53$ min and $55$ min after $T_0$, respectively. The \textit{Fermi}-GBM data shows a continuous prompt emission lasting more than 600 seconds in the 8 keV – 40\,MeV energy range. The \textit{Swift} BAT and XRT instruments identified an X-ray extremely bright afterglow that is one order of magnitude brighter at $T_0 + 4.5$\,ks than any previous GRB observed by \textit{Swift} \citep{221009A_swift}. The event saturated almost all the gamma-ray detectors due to its exceptional high fluence.}

{A multi-wavelength extensive campaign was performed covering the entire electromagnetic spectrum \citep{mwl_221009A}. The redshift was estimated to be $z = 0.151$ \citep{2022GCN.32648....1D}. In the VHE domain, the LHAASO experiment observed the event starting from the trigger time up to $\sim 6000$ s \citep{lhaaso23}. A very clear signal with a statistical significance $ > 250$ standard deviation was identified. The temporal evolution exhibits a smooth profile with a rapid rise to a peak, followed by a decay lasting for $3000 $ s after the peak. The intrinsic flux spectrum in the energy range from $\sim 200$ GeV to $\sim 7$ TeV was fitted in several time intervals during the rise and decay phase with a power law with no evidence of spectral breaks or cut-offs up to $5$ TeV. Further studies published in \citet{GRB_lhaaso_13TeV} indicates the evidence for a rising cutoff at energies of $\sim 13$ TeV. The intrinsic spectral photon index was estimated in the range of $\approx 2.1 - 2.5$ (depending on the EBL model and the time interval considered) showing a mild spectral hardening in time. The smoothed temporal behaviour without highly variable structures, the absence of spectral cut-off or breaks, and the modest spectral evolution indicate that the VHE $\gamma$-ray component seen by LHAASO belongs to the GRB afterglow phase.}

{The highly exceptional nature of this GRB has been studied considering the population of GRBs detected thus far. It has been estimated that in terms of energetics, fluence, and distance, such events occur at a rate of $\sim 1$ per $1000- 10000$ yr \citep{221009A_boat,221009A_swift}, which makes this event a remarkable object unlikely to be repeated for long time.
}

\subsection{Model of the intrinsic emissions and simulation of cascade emissions of GRB\,190114C and GRB\,221009A}
Among the GRBs, GRB\,190114C is the first one detected at TeV energies in the first $40$ minutes of observations \citep{miceli_nava}. As a result, the VHE spectral properties and time evolution have been studied with unprecedented detail. For this reason we decide to use the spectral and temporal properties of this event as a proxy to study the expected time-delayed secondary emitted cascade from a generic GRB. We assume the average flux emitted from the source for the time activity limited by TeV detection and estimate the secondary cascade emission through CRPropa simulation \citep{crpropa3}.

The calculation of the average emitted flux $\langle F \rangle_{6-2454}$ covering the entire afterglow time interval $6-2454$\,s is estimated as:
\begin{equation}
    \langle F \rangle_{6-2454} = \frac{\int_6^{2454} F(t) dt}{\Delta t}
,\end{equation}
where $F(t)$ is the time-evolving flux extrapolated from MAGIC results from 6\,s to 2454\,s in the $0.2-10$ TeV energy range. For this estimate, we can exploit the results of the modeling of GRB\,190114C presented in \citep{magic_14C_modeling}. Here, the intrinsic differential photon spectrum in the time interval $68-180$\,s is described with a log-parabola function in the energy range $0.2-10$ TeV (as also shown in \citet{paolo_paper}):
\begin{equation}
\bigg(\frac{dN}{dE}\bigg)_{int} \propto \bigg(\frac{E}{E_0}\bigg)^{-\Gamma- 0.2\log{E/E_0}}
\label{eq:intrinsic_spec}
,\end{equation}
where the spectral index $\Gamma = 2.5$. This modelization is robust both from the observational and theoretical side. Indeed, it reproduces the MAGIC data and also takes into account the modification in the VHE tail spectrum expected from theory and due to Klein-Nishina effects and $\gamma$-$\gamma$ absorption. This assumption implies that the spectral evolution is driven only by the normalization factor, and there is no variation of the spectral shape over time. This is in agreement with MAGIC results (\cite{magic_14C_modeling}, see section on methods), although an hint of a possible spectral evolution to a softer value should not be excluded. For the time evolution of the flux, we assume a power-law temporal decay with index $-1.6,$ as calculated by the MAGIC Collaboration in the $62-2454$\,s time interval. We consider  the afterglow phase to have started at
$6$\,s, as supported by \citep{ravasio_190114C}.

{GRB\,221009A is the brightest GRB ever detected, with a clear identification of an emission component in the GeV-TeV domain as well. Its extraordinary intrinsic nature, in terms of energetics, luminosity, and distance, makes it the most promising transient target for IGMF studies so far. In the following, we propose an estimate for the expected pair echo emission for GRB\,221009A. In addition, the properties of this exceptional bright source have been used to test how the choice of the VHE intrinsic spectrum affects the shape and amount of the expected pair echo emission. The LHAASO Collaboration reported that the intrinsic spectrum can be represented with a power law spectrum that also extends into the TeV domain. For this reason, we followed a different approach with respect to GRB\,190114C: considering the table S2 in \citet{lhaaso23} (supplementary materials), we evaluated the time-weighted average intrinsic spectrum in the 5 time bins reported between 231 and 2000 s after the trigger burst. Since in \citet{GRB_lhaaso_13TeV} the authors reported a maximum energy of 13\,TeV, we finally add an exponential cutoff to the power-law spectrum at 13\,TeV. Therefore, for the GRB\,221009A case, the final intrinsic VHE spectrum chosen for the simulations is:}

\begin{equation}
\bigg(\frac{dN}{dE}\bigg)_{int} \propto \bigg(\frac{E}{E_0}\bigg)^{-\Gamma}\times\exp{(-E/E_{cut}),}
\label{eq:intrinsic_specGRB221009A}
\end{equation}

\noindent {where E$_{0}$=1\,TeV, $\Gamma$=2.3 and E$_{cut}$=13\,TeV.}

With this choice of the intrinsic differential photon {spectra,} we used CRPropa to simulate the pair echo emission for different IGMF settings and exposure times. CRPropa is a Monte Carlo code that traces the development of the cascade emission produced by the interaction of VHE photons with the intergalactic medium. The source is located at the center of the sphere with a radius equal to the source distance. {For the case of GRB\,190114C,} we assume that the source emits VHE photons between 0.2 and 10 TeV within a cone of 10$^\circ$ degrees of aperture, {which can be considered as a standard GRB jet aperture}.  {Concerning GRB\,221009A, we inject VHE photons in the range 0.1-30 TeV. Since for this GRB in \cite{lhaaso23} the authors estimated the GRB jet aperture of $\sim$1.6$^{\circ}$, we assumed this value as emission cone aperture in our simulations.} During propagation in the intergalactic medium, the VHE photons interact via $\gamma$-$\gamma$ pair production with the EBL and then the pairs lose energy via inverse Compton with the CMB. Concerning the target photon field for pair production we use the \citet{franceschini08} EBL model. A cascade photon that hits the sphere and has an energy larger than 10 GeV represents a particle arriving and being detected at Earth. The magnetic field is assumed to be a turbulent zero-mean Gaussian random field with a Kolmogorov spectrum; it is defined in Fourier space, transformed into real space, and then projected onto a uniformly spaced cubic grid with $N=100^3$ cells of 50 Mpc size. The minimum and maximum scale lengths in which the magnetic field is defined are 1 Mpc and 25 Mpc, respectively. In this configuration, the correlation length is about 5 Mpc. The cubic cell is then periodically repeated to cover the distance between the source and Earth. For each magnetic field strength tested (root mean square), we injected $10^3$ primary VHE photons distributed as in Eq. (\ref{eq:intrinsic_spec}) {for GRB\,190114C} and {Eq. (\ref{eq:intrinsic_specGRB221009A}) for GRB\,221009A} repeat the simulation for 200 times (in total $2\times10^5$ primary VHE photons). In each run, we change the seed used to generate the magnetic field grid to avoid spurious features due to the choice of that particular realization of a magnetic field. All particles are traced with a minimum step size of 10$^{-6}$ pc, which is sufficient to reproduce time delays, with an accuracy better than 2 minutes. To produce the pair echo SEDs and light curves we use the same procedure described in \cite{paolo_paper}: assuming the observer perfectly aligns with the source cone axis, we computed the cascade spectrum within a certain observation time bin $\Delta T$ using the Eq. (4) in \cite{paolo_paper} and perform the time average over different exposure times to take into account the time dilution of the signal.

\begin{figure}
    \includegraphics[width=0.5\textwidth]{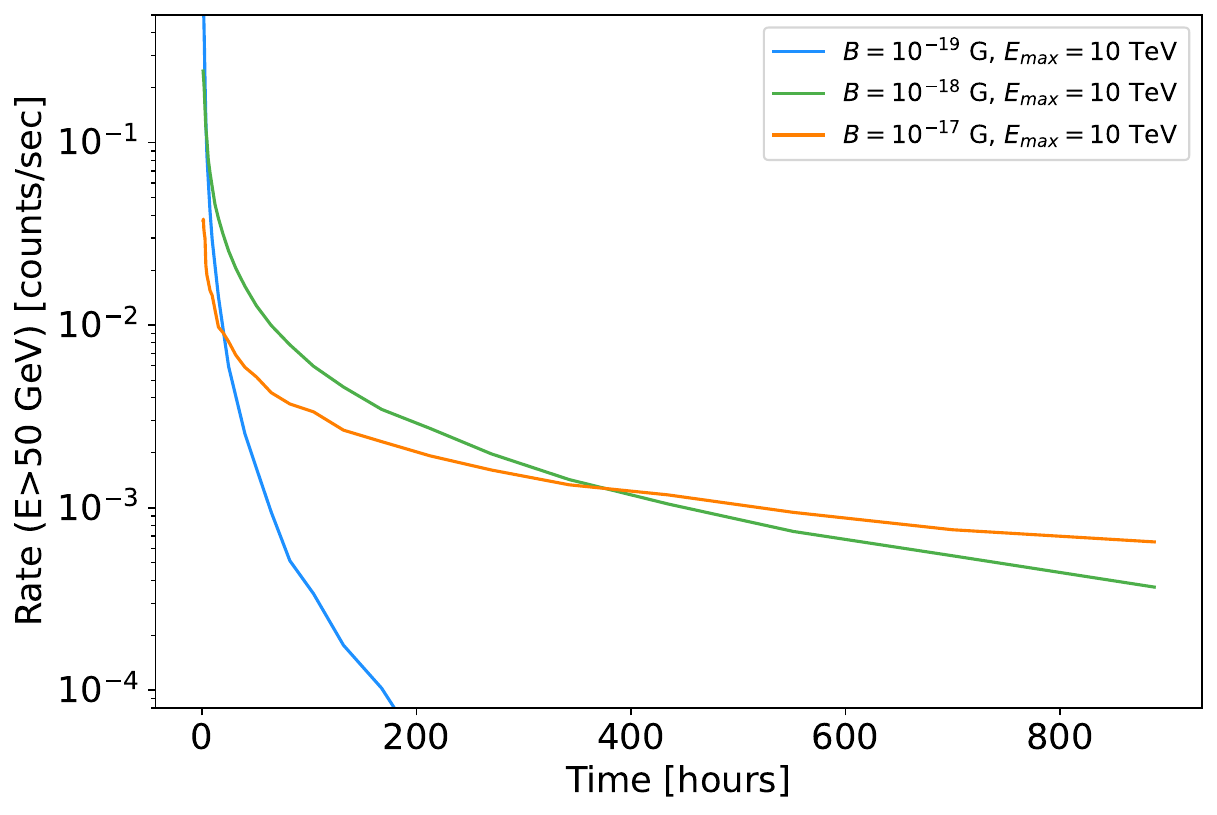}
   \caption{Simulated count-rate light curve of the pair echo emission for $E > 50$ GeV for three different values of IGMF strenghts and a maximum energy of 10 TeV.}
\label{fig:lc_count_rates}

\end{figure}

An example of pair echo emission derived with this approach is shown in Fig.~\ref{fig:lc_count_rates}, showing the count rate above $50$ GeV after the afterglow onset. Here, we assume three strengths of the magnetic field {($B = 10^{-19}$ G, $B = 10^{-18}$ G, and $B = 10^{-17}$ G)} and a maximum energy for the intrinsic photon spectrum (see Eq. \ref{eq:intrinsic_spec}) of $E_{max} = 10$ TeV. The bulk of the echo emission is expected to arrive in the first hours after the afterglow onset. As expected, stronger IGMF strengths result in more diluted cascade emission, since pairs experience a longer delay.

\section{Results}
\label{sec:results}
We exploit the phenomenological model{s} described in Sect. \ref{sec:methods} to optimize the observational strategy for VHE detectors to identify the pair echo emission coming from GRBs. For this study, we focus on the performances of the MAGIC and CTA-North array. As a method, we compared the expected secondary cascade emission from GRBs with the flux sensitivity of the MAGIC telescopes and the CTA-North array. Such a comparison can give an indication of the possible IGMF signatures that can be probed with current and future generations of VHE instruments. We assume different observational time windows and identify the most promising ones. Moreover, we investigate the impact of different GRB intrinsic properties on the detection capabilities. 

\begin{figure*}
    \includegraphics[keepaspectratio=false,width=6.5cm,height=5cm]{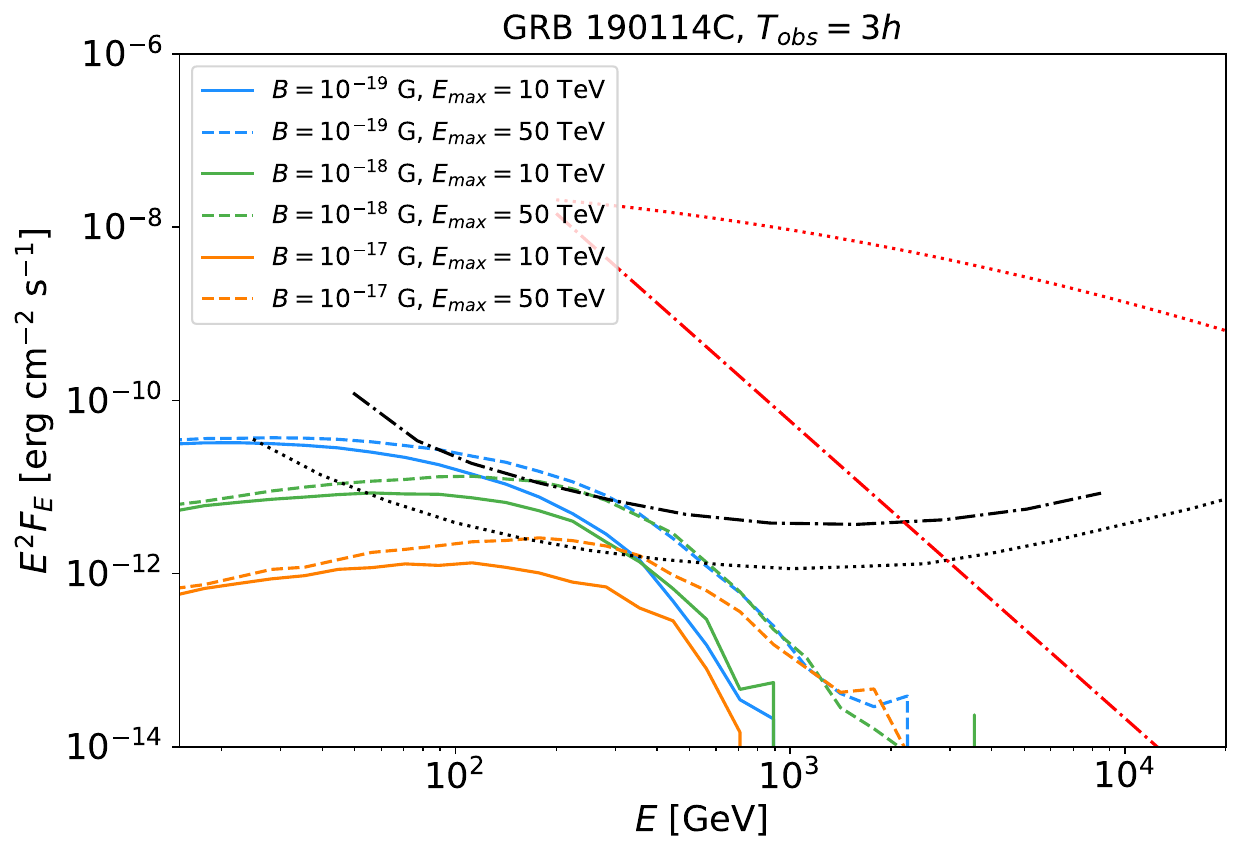}
    \includegraphics[keepaspectratio=false,width=5.5cm,height=5cm]{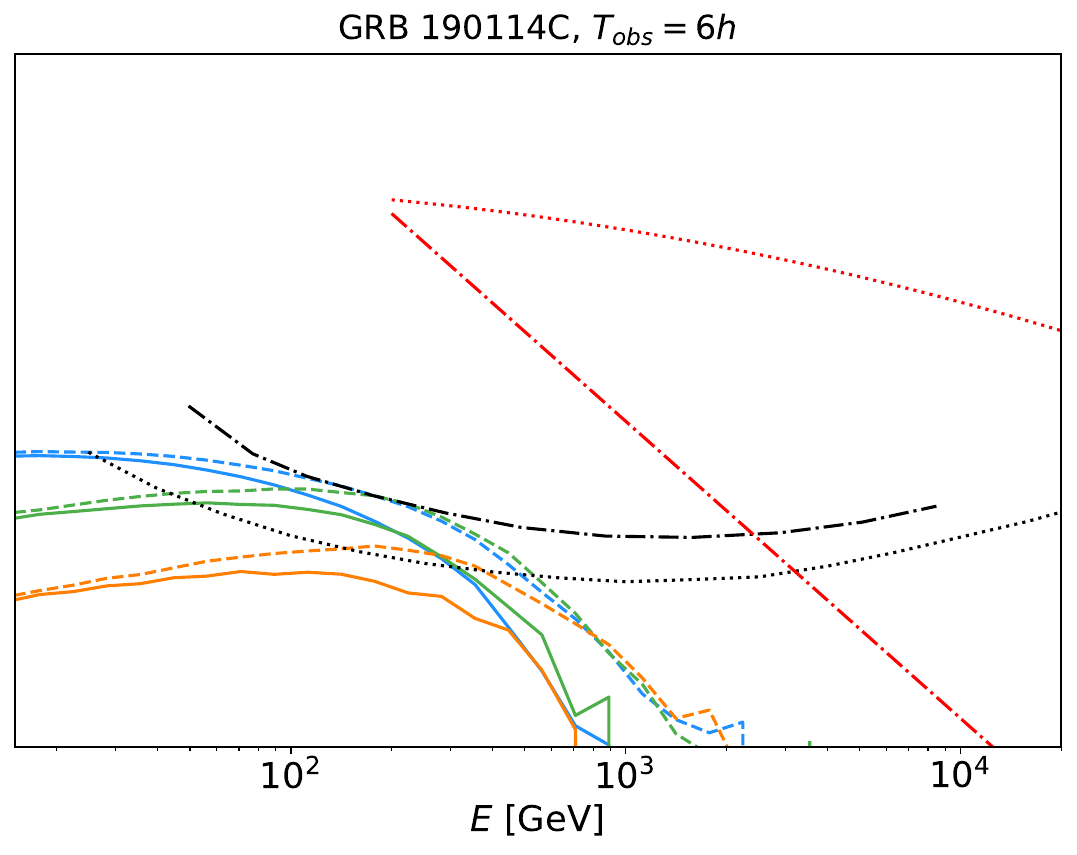}
    \includegraphics[keepaspectratio=false,width=5.5cm,height=5cm]
    {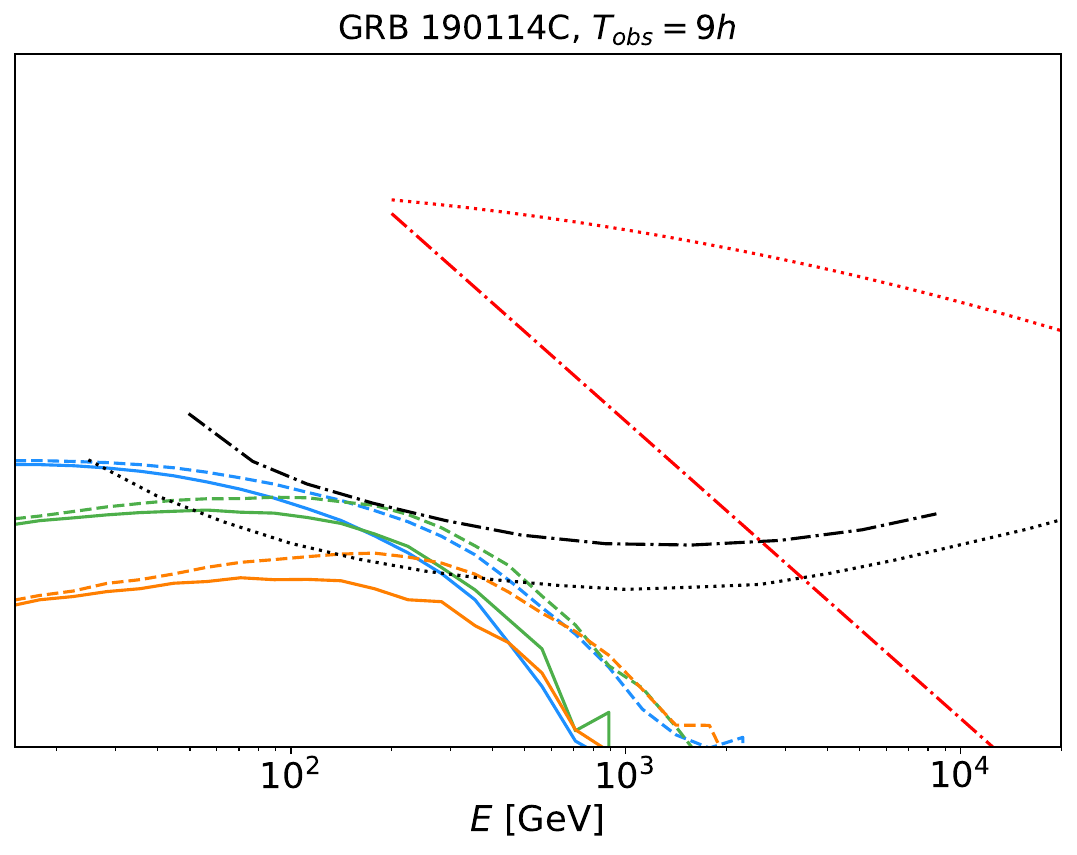}
    \\
    \includegraphics[keepaspectratio=false,width=6.5cm,height=5cm]{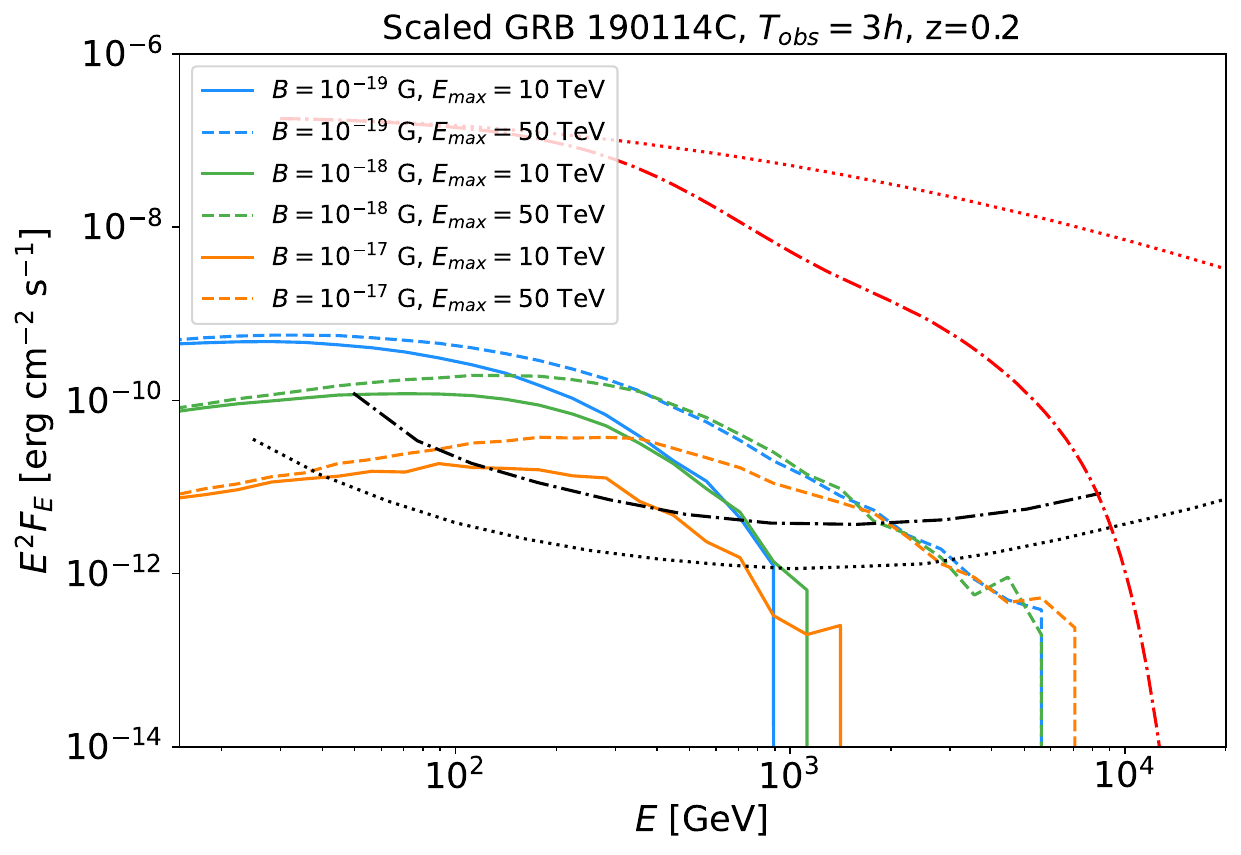}
    \includegraphics[keepaspectratio=false,width=5.5cm,height=5cm]{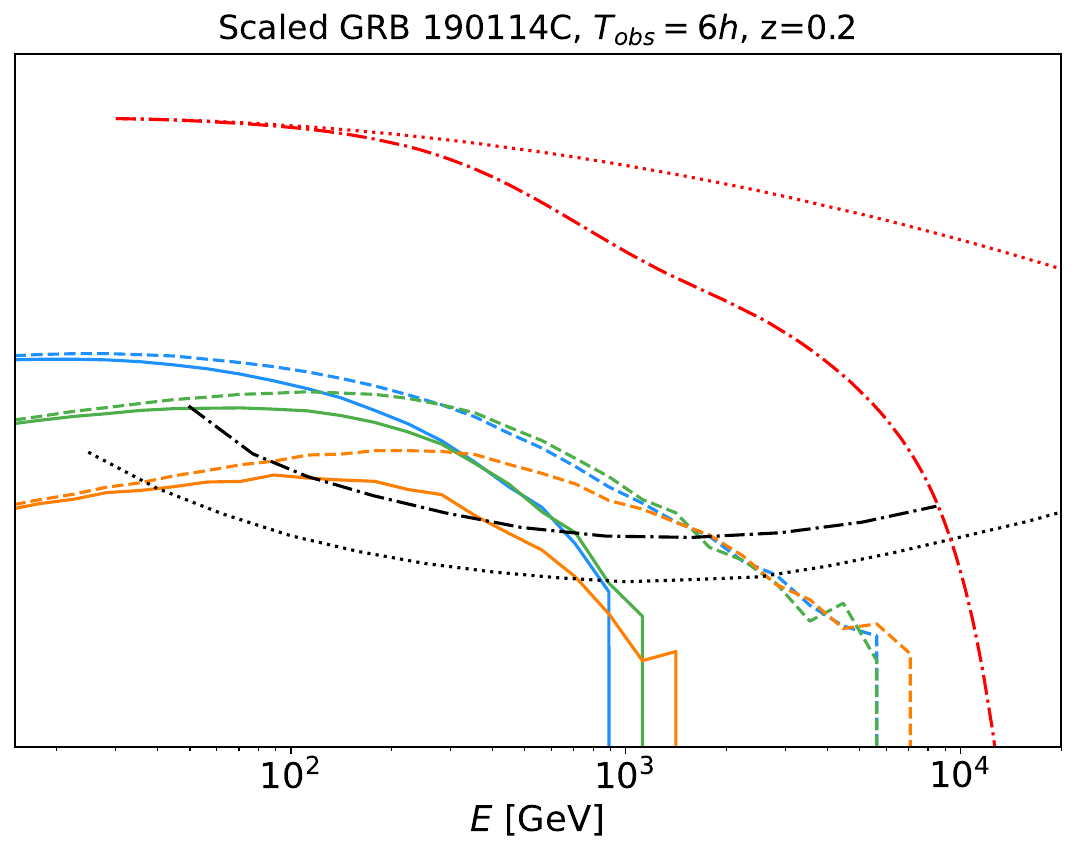}
    \includegraphics[keepaspectratio=false,width=5.5cm,height=5cm]{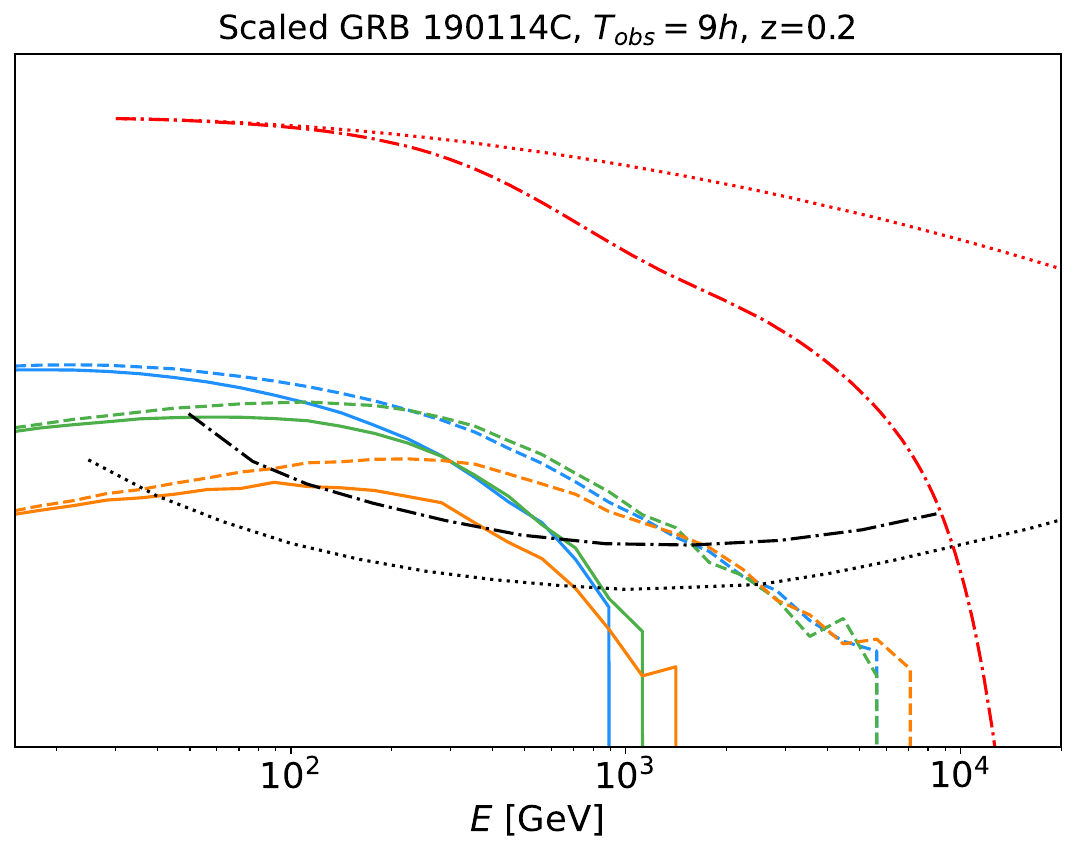} \\
    
      \includegraphics[keepaspectratio=false,width=6.5cm,height=5cm]{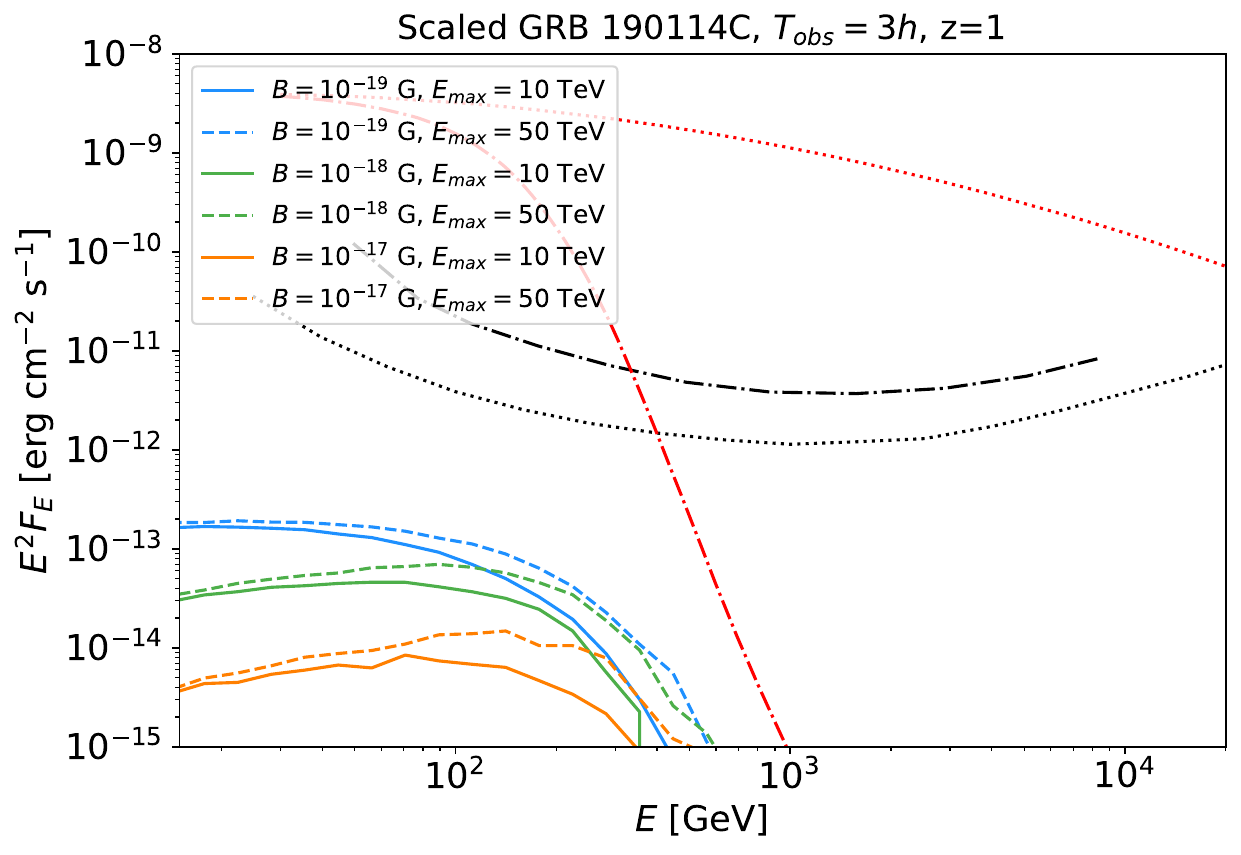}
    \includegraphics[keepaspectratio=false,width=5.5cm,height=5cm]{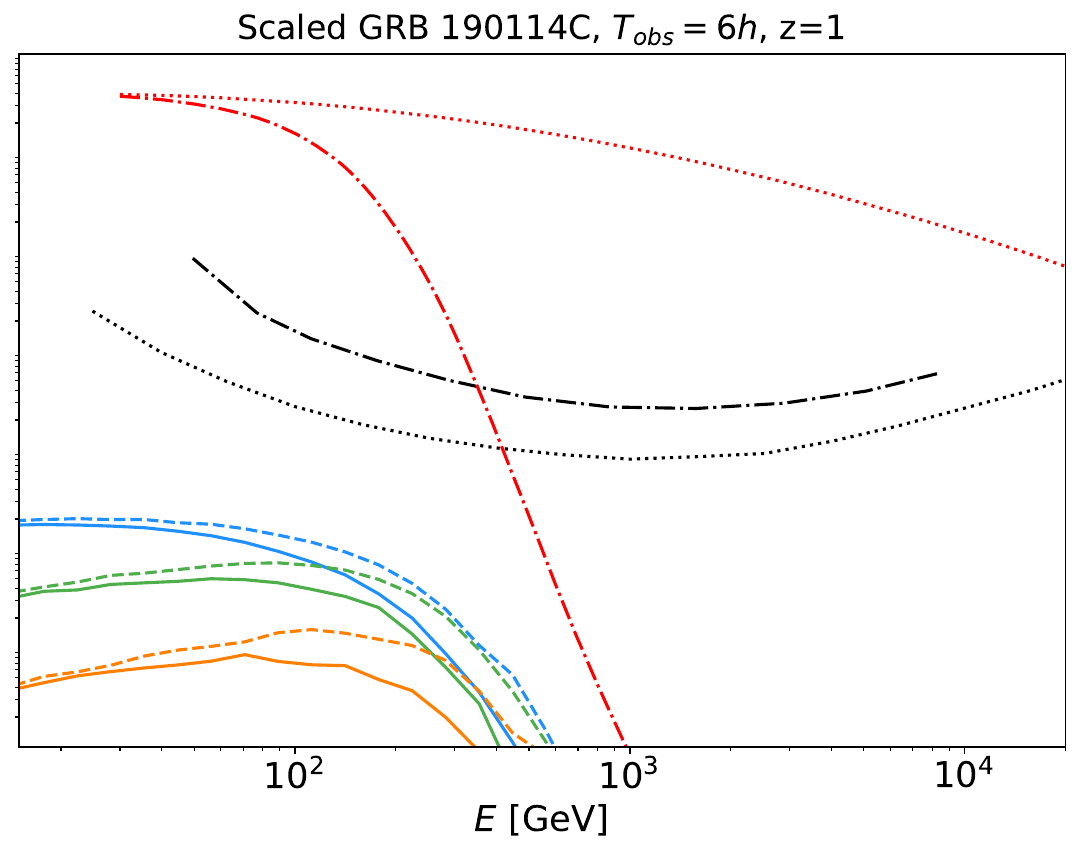}
    \includegraphics[keepaspectratio=false,width=5.5cm,height=5cm]{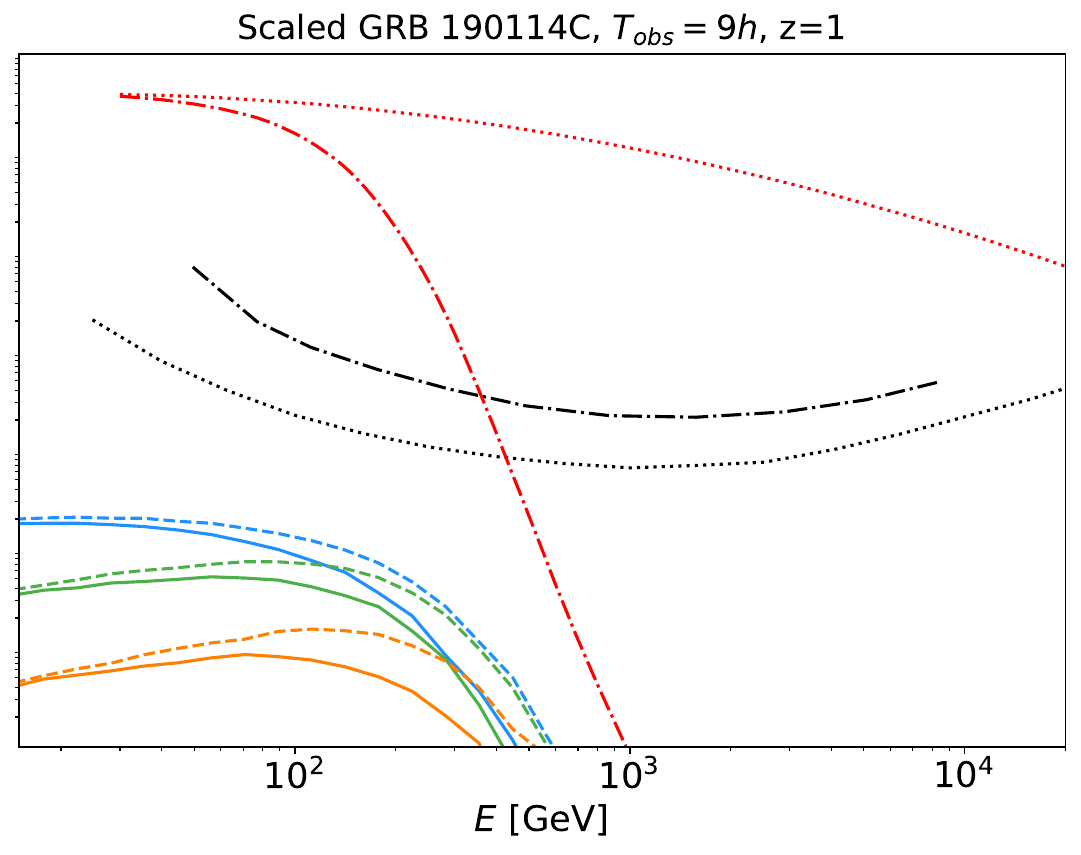}
    \caption{Comparison between the time-averaged SEDs of the secondary cascade output (pair echo emission, color solid, or dashed lines) and the MAGIC (black dash-dotted lines) and the CTA-North (black dotted lines) sensitivity curves for three different observational times (respectively from left to right: 3\;h, 6\;h, 9\;h). Three different cases have been considered: GRB\,190114C, with a redshift of $z=0.42$ ({upper panel}), a scaled GRB\,190114C with  $z = 0.2$ ({middle panel}), and a scaled GRB\,190114C with  $z = 1.0$ ({bottom panel}). The GRB intrinsic (red dotted lines) and observed (red dash-dotted lines) spectra are also shown. Values of the IGMF strengths and maximum energy for primary photons considered are shown in the legend.} 
    \label{fig:plot_grbs}
\end{figure*}

First, we considered the case of GRB\,190114C (z = 0.42). We started by estimating the secondary photon cascade simulated with CRPropa from $t = 3000$\,s after $T_0$, which approximately marks the end of the TeV detected afterglow. This is a safe approach, analogous to previous studies, used to exclude possible contamination of the primary emission coming from the source. Then, we estimated the time-averaged SEDs of the pair echo emission {assuming three IGMF strength values ($10^{-19}$\;G, $10^{-18}$\;G, and $10^{-17}$\;G)} and two values of the maximum energy for the intrinsic source spectrum ($10$\;TeV and $50$\;TeV). As time observational windows, we consider three possible durations, compatible with typical observational time windows of transient objects with IACTs during the night: $3$ hrs, $6$ hrs, and $9$ hrs (Fig.~\ref{fig:plot_grbs}, upper panel, respectively from left to right).

\begin{figure*}
    \includegraphics[keepaspectratio=false,width=6.5cm,height=5cm]{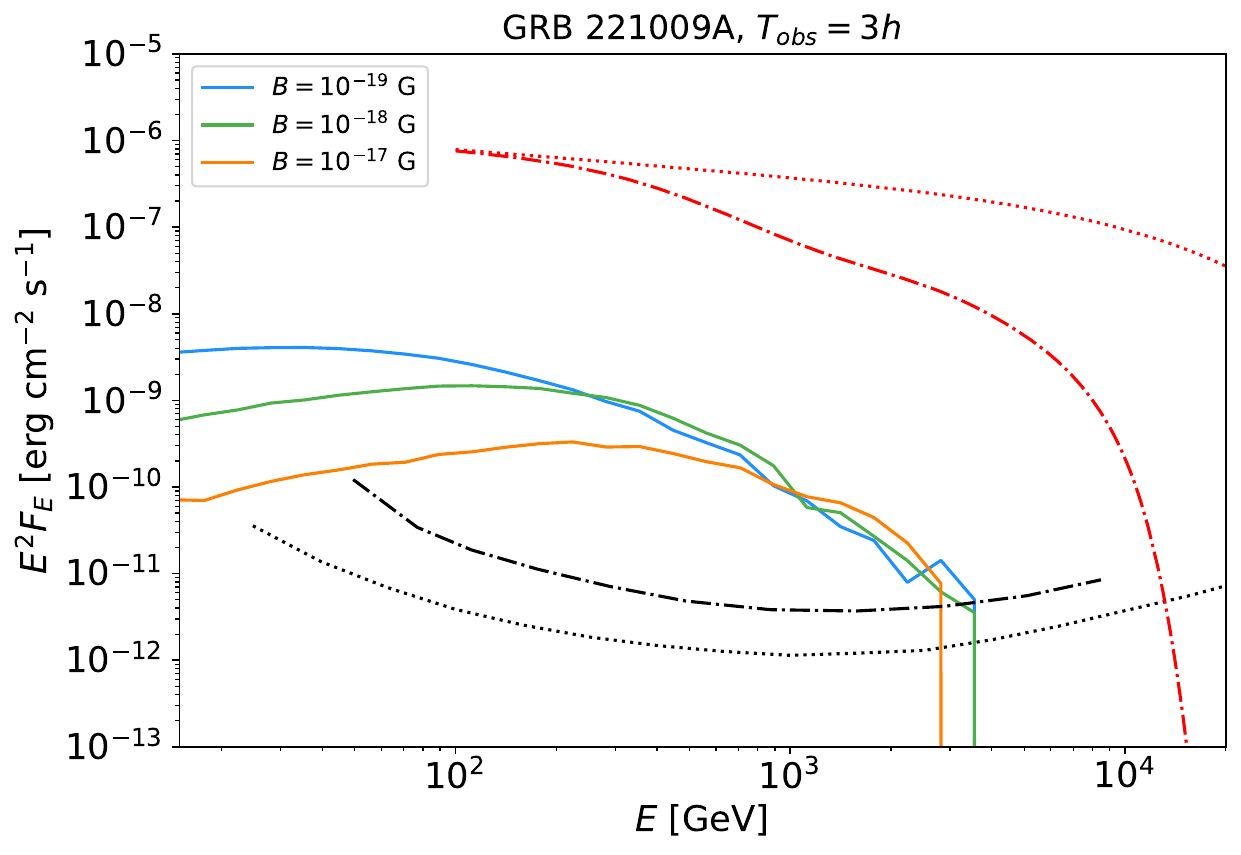}
    \includegraphics[keepaspectratio=false,width=5.5cm,height=5cm]{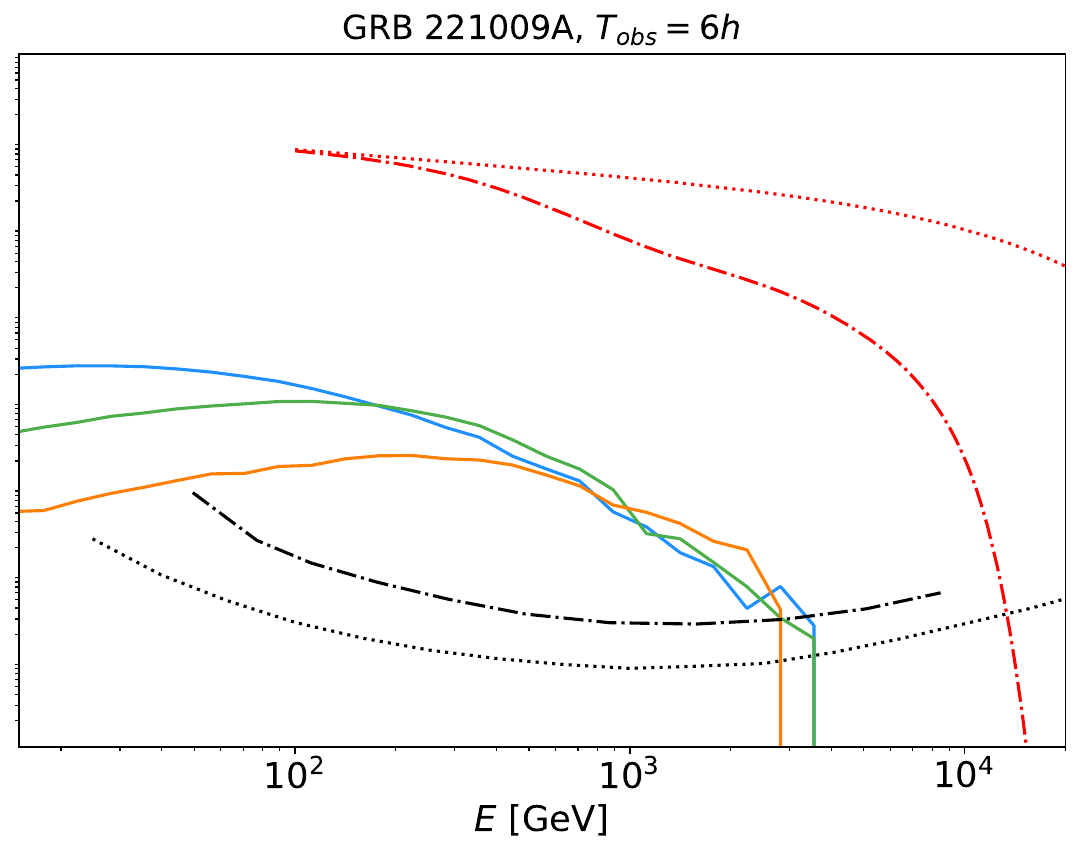}
    \includegraphics[keepaspectratio=false,width=5.5cm,height=5cm]{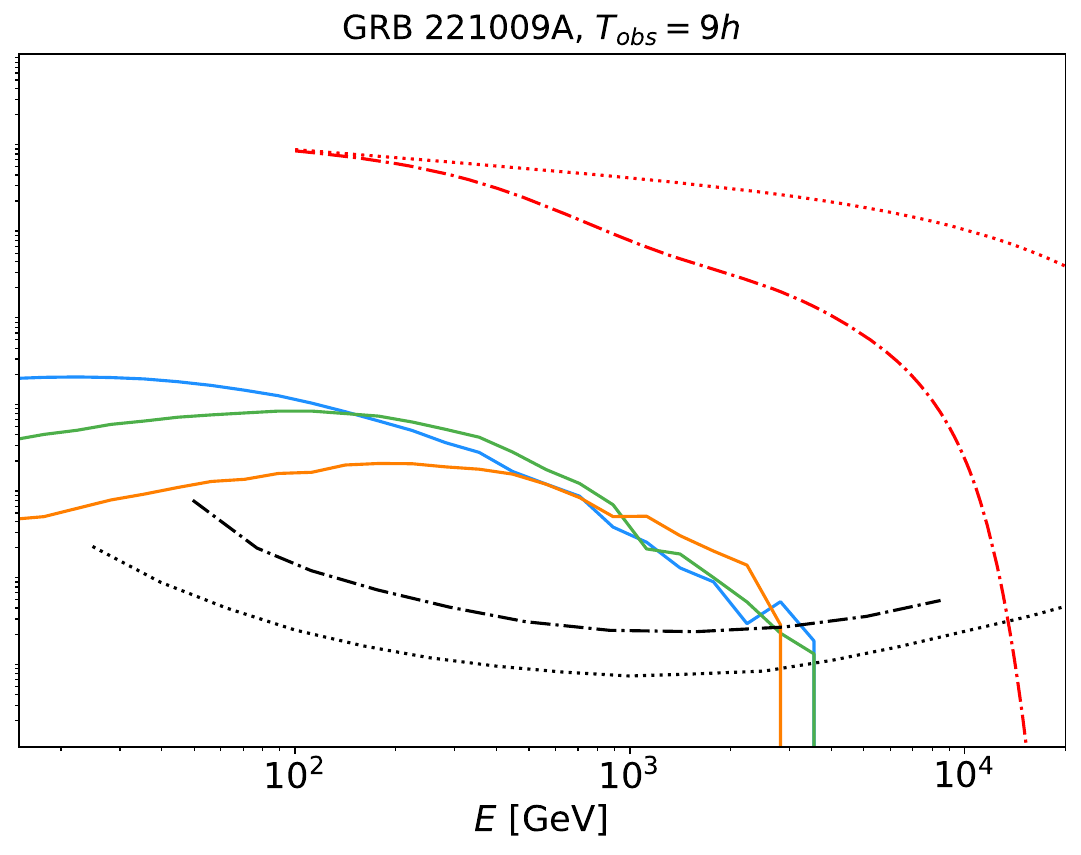}
    \\
    \includegraphics[keepaspectratio=false,width=6.5cm,height=5cm]{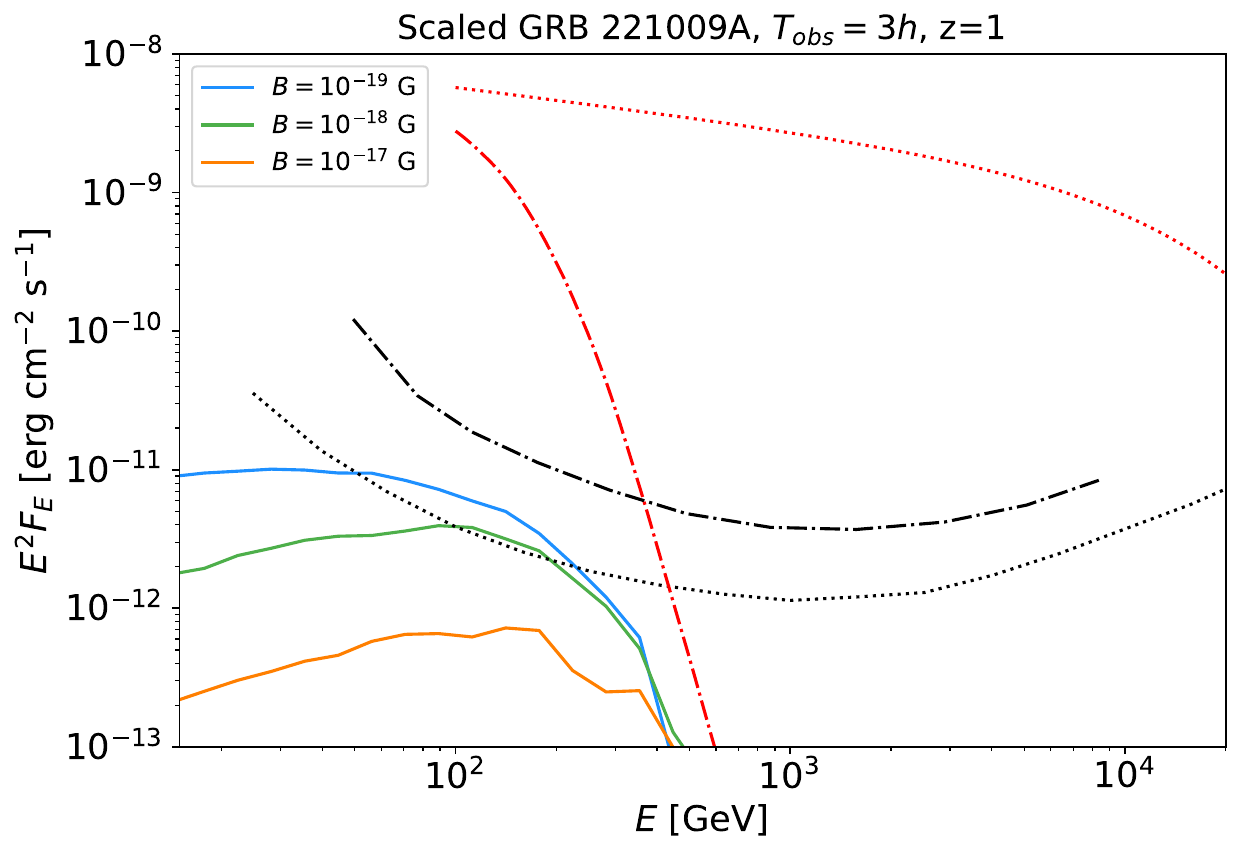}
    \includegraphics[keepaspectratio=false,width=5.5cm,height=5cm]{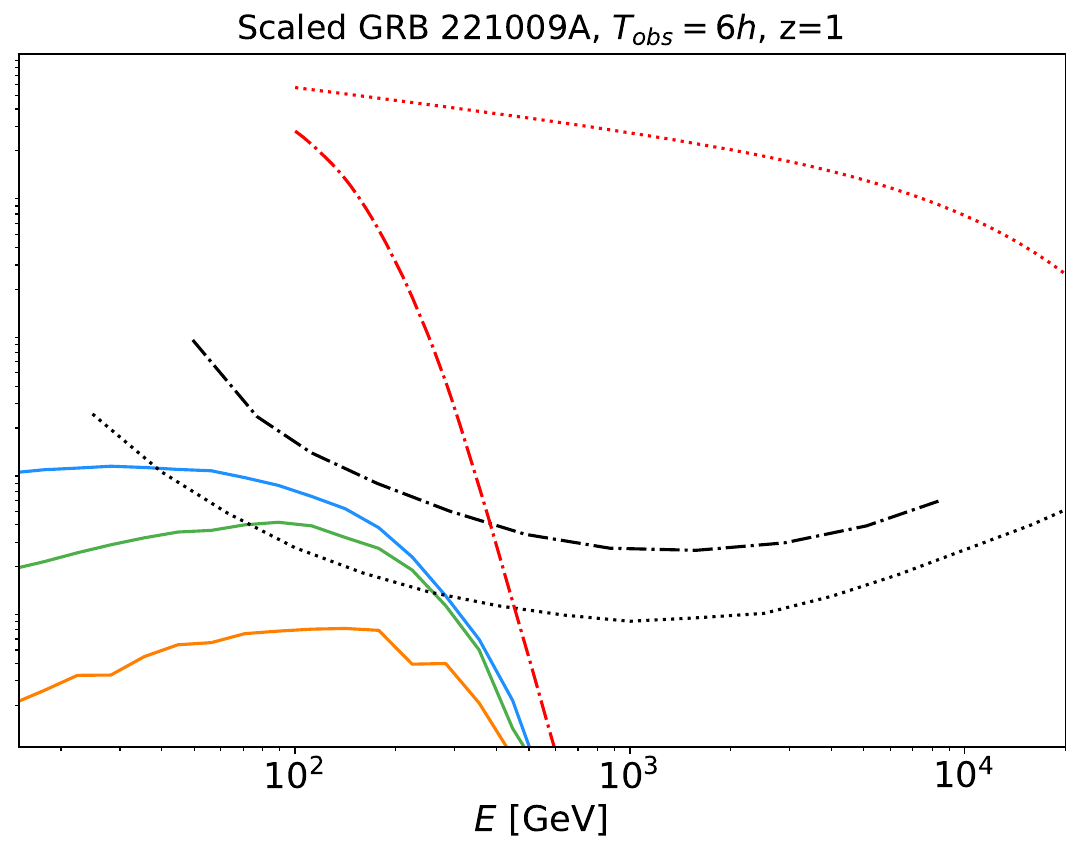}
    \includegraphics[keepaspectratio=false,width=5.5cm,height=5cm]{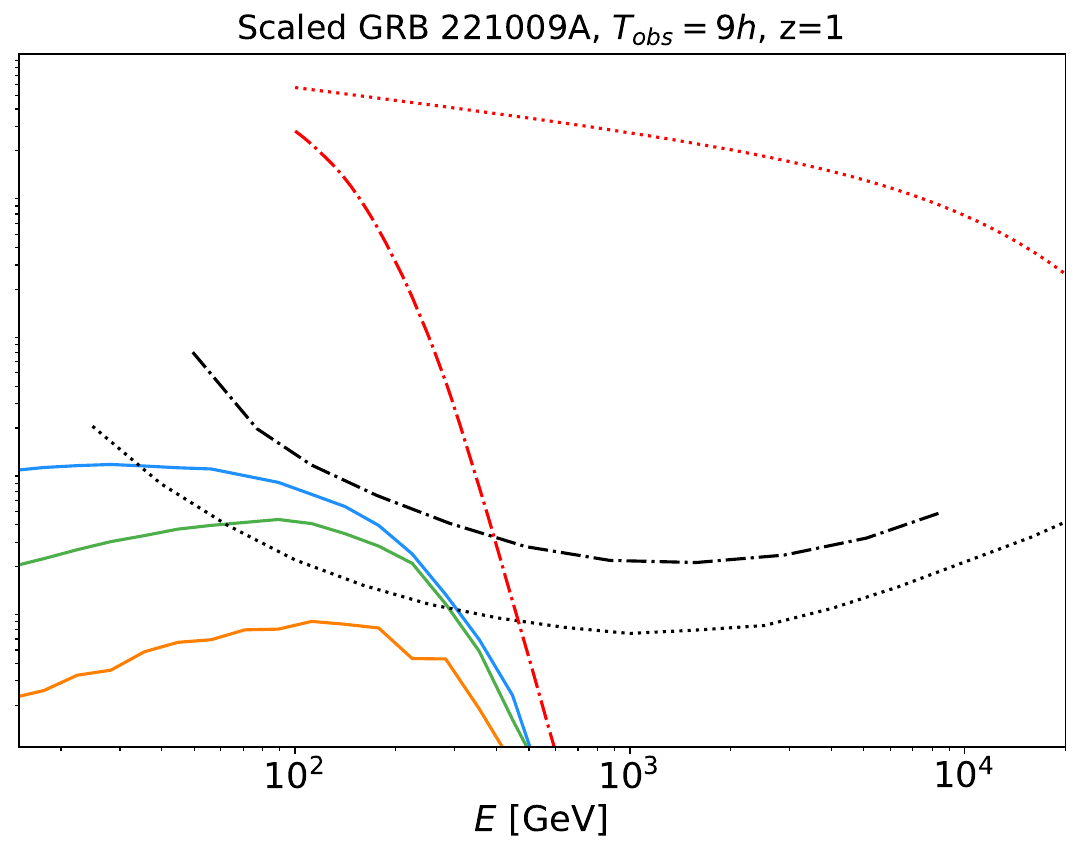} \\
    
    \caption{Comparison between the time-averaged SEDs of the secondary cascade output (pair echo emission, color solid lines) and the MAGIC (black dash-dotted lines) and the CTA-North (black dotted lines) sensitivity curves for three different observational times (respectively from left to right: 3\;h, 6\;h, 9\;h). Two different cases are shown: GRB\,221009A, $z=0.15$ ({upper panel}) and a scaled GRB\,221009A with redshift $z = 1$ ({bottom panel})
    }
    \label{fig:plot_grbs221009A}
\end{figure*}

{Second, we considered the more extreme case of GRB\, 221009A (z = 0.15). As reported in \cite{lhaaso23} the time window in which the VHE emission has been detected is 3000 s. For this reason, for this case we also considered cascade photons arrived only from $t = 3000$\,s after $T_0$ to avoid possible contamination due to the presence of the very bright afterglow. As for GRB\,190114C, we computed the expected cascade SEDs for different exposure times and magnetic field strengths (Fig.~\ref{fig:plot_grbs221009A}, upper panel)}.

The sensitivity for a $5 \sigma$ significant detection of the MAGIC and CTA-North telescopes has been derived from \footnote{\url{https://www.cta-observatory.org/science/ctao-performance/\#1472563157332-1ef9e83d-426c}}. The $50$ hours differential sensitivity curves have been rescaled to the corresponding observational time windows adopted in this study ($3$ hrs, $6$ hrs, and $9$ hrs) assuming that the sensitivity, $S,$ scales with time $t$ as $S \propto 1/\sqrt{t}$. This is a simplified approach since it does not take into account the variation of the instrument response functions of the telescopes (differential flux sensitivities, effective areas, angular or energy resolutions) with time. However, these results are in agreement within a factor of 2 with the one obtained in \citep{fioretti}, where specific estimates have been derived for transient source observations at different observational times. 
The sensitivities of the \textit{Fermi}-LAT instrument have been also derived for the same observational windows following an analogous procedure. However, we decided not to show them in this study, since the estimated sensitivities are about two to four orders of magnitude higher than the one calculated for the MAGIC and CTA telescopes. The result shows that for such short delays \textit{Fermi}-LAT cannot provide any relevant constraints on the IGMF strength.

In the considered case of GRB\,190114C, the secondary cascade output flux is above all three (t $>$ 3\;h, 6\;h, 9\;h) CTA-North sensitivity curves from few tens of GeV up to a few hundreds of GeV {(with the only exception of the configuration with $B = 10^{-17}$ G and $E_{max} = 10$ TeV)}, while it is marginally below {or at the same level as the MAGIC ones for weak field of $B = 10^{-19}$ G}. This implies that such component, if present, would have been clearly detected by CTA-North or, in case of non-detection, the magnetic field configurations would have been excluded and a lower limit could have been derived. For the MAGIC telescopes, the discrimination power is lower because of their weaker sensitivity and this component would have been only marginally detected or not detected at all {in the case of GRB\,190114C}. Moreover, it is notable that extending from $3$ hrs to $9$ hrs does not increase the detection chances for the telescopes. This is because for such weak IGMF strengths, the reprocessed emission above $\sim 50$ GeV is emitted mostly in the first hours after the afterglow emission and the dilution with time is not so strong. Larger IGMF strengths increase the dilution of the emission in time, but at the price of a lower cascade flux in the first hours, which reduces the chances of detection.

{Similar considerations are also valid for the case of GRB\,221009A, where it can be seen that the expected output cascades for the IGMF configurations considered in this study are well above the MAGIC and CTA-North sensitivities. This indicates that even stronger magnetic fields $B > 10^{-17}$\,G can potentially be constrained and studied in this case. However, it should be noted that due to the exceptional properties of this event, the results on this GRB are not representative of the population of GRBs at VHE, but should be instead considered as a very optimistic case.} 

The approach presented here and the developed phenomenological model{s} could be extended to other GRBs which may display different intrinsic properties. One of the main features (which also affects the secondary cascade emission) is the cosmological distance of the source, which significantly changes the EBL density along the line of sight and the reprocessed emission that is derived. To test the effect of distance on the emitted cascade power, we simulated  {three} events: {i) two } assuming that they have the same intrinsic rest-frame luminosity and intrinsic spectral shape of GRB\,190114C, but assuming a more nearby source with redshift $z = 0.2$ and a more distant source with redshift $z = 1.0$; {ii) and one  assuming the same intrinsic rest-frame luminosity and intrinsic spectral shape of GRB\,221009A but assuming a more distant object ($z=1$). In Fig.~\ref{fig:plot_grbs} we refer to the former two simulated events as scaled GRB\,190114C, while in Fig.~\ref{fig:plot_grbs221009A} we refer to the latter case  as scaled GRB\,221009A}. These redshift values are in agreement with the population of GRBs detected to date at TeV energies \citep{miceli_nava}. We simulated the expected cascade output in the same observational time window and the same magnetic field configurations as explored {previously} and we compare it again with the MAGIC and CTA-North sensitivities {scaled GRB\,190114C: Fig.~\ref{fig:plot_grbs}, middle panel for  $z=0.2$, respectively, from left to right; Fig.~\ref{fig:plot_grbs}, bottom panel for  $z=1.0$, respectively, from left to right; \textit{Scaled GRB\,221009A}: Fig.~\ref{fig:plot_grbs221009A} bottom panel for  $z=1.0$, respectively, from left to right).} For the scaling of GRB\,190114C, the results from the comparison of the secondary output cascade and the MAGIC and CTA-North sensitivities show that for the GRB simulated at redshift $z=1.0$, the pair echo emission is below the MAGIC and CTA-North sensitivity of more than two orders of magnitude. On the other hand, for the nearby GRB, at $z = 0.2$ the pair echo emission is well above both MAGIC and CTA-North sensitivities.
{For the scaling of GRB\,221009A, it can be seen that at a redshift of $z = 1.0$, the pair echo emission is below the MAGIC sensitivities in all the configurations studied; however, it is still visible for CTA-North for the weaker magnetic fields cases ($B = 10^{-19}$ G and $B = 10^{-18}$ G). The increase of the observational time window also slightly increase the chances of detection.}

\section{Conclusions}
\label{sec:conclusion}
In this paper, we explore the capabilities of current and future VHE detectors for IGMF studies. We estimate the pair echo emission generated by GRBs by means of the simulation code CRPropa. For this purpose, we built phenomenological model{s} for GRB emission in the VHE domain, based on the results and interpretation of GRB\,190114C {and GRB\,221009A}. We then calculated the time-average SEDs of the pair echo emission for different IGMF strengths and different GRB intrinsic properties and observational conditions. Finally, we compared these spectra with the sensitivity of the MAGIC telescopes and the CTA-North array in the GeV-TeV domain. 

Thus far, the secondary emission cascade has always been estimated in a lower energy band, between $\sim 0.1-10$ GeV exploiting long-time exposures of satellite instruments. Indeed, in this energy range space satellites such as the \textit{Fermi} telescope can provide good sensitivity and can collect data continuously for weeks, months or even years, while IACTs cannot provide any results since their energy threshold is typically around $\sim 50-100$ GeV for the current instruments or around $\sim 20-50$ GeV for the future generation ones and their observational times are on the order of several hours. 

We focused our study on short time delays (a few hours after the burst trigger) and magnetic field {strengths in the range $10^{-17}$-$10^{-19}$ G).} This choice is driven by theoretical considerations since the time delays, $t_{delay}$, induced by IGMF to the pair-echo emission follow the relation, $t_{delay} \propto B^2 E_{rep}^{-5/2}$, for long correlation lengths and $t_{delay} \propto B^2 E_{rep}^{-2}$ for short correlation lengths \citep{neronov09}; here, $B$ is the assumed IGMF strength and $E_{rep}$ is the energy of the pair-echo emission. In the energy range where IACTs are sensitive, from few tens/hundreds of GeV up to a few tens of TeV, the delay is shorter than the one experienced by lower energy photons. These delays are compatible with the IACTs observational capabilities and the study of "weak"\ IGMF values, ranging between $\sim 10^{-17}$\;G -- $10^{-21}$\;G. Larger IGMF strengths will increase the delay of the VHE photons and generate a more diluted cascade in time which will make difficult for IACTs to be sensitive to the pair-echo emission. 

It is interesting to note that at these time intervals, {the delay due to the angular spread of the electron-positron pairs in the gamma-gamma pair production can become relevant. In such a case, this delay can be concurrent with the one induced by the IGMF and should be considered in the calculation of the delayed emission}. The impact of the former effect is typically not taken into account in numerical simulations. From \citet{neronov09}, we derived that the time delays after which this effect can be considered negligible for a $1$\,GeV and a $100$ GeV photon are $\sim 6.2$\,hrs and $\sim 22$ s, respectively. {Considering the energy range of the Cherenkov telescopes treated in this study, we can conclude that this effect is negligible with respect to the one induced by the IGMF; however, it should be considered for the studies of satellite instruments working in the sub-GeV - GeV domain}. In conclusion, longer delays and deep exposures of weeks, months, or years are convenient for satellite instruments working at energies below a few tens of GeV. On the other hand, shorter delays can be explored and are most suitable for the VHE band and the IACT observational abilities.

In this paper, we demonstrate that VHE detectors can also provide relevant information for studying pair echo emission and therefore test IGMF configurations {with values that are competitive with the current most stringent constraints derived from AGNs. This may also offer the opportunity to have an independent verification of such results using completely different sources.} This is a scientifically relevant case also for the CTAO southern array in light of the enhancement proposed with the CTA+ project(\footnote{\url{https://pnrr.inaf.it/progetto-ctaplus/}}).
Typically, transient sources are observed by IACTs starting from $\sim$ 30 s up to a few hours after the trigger burst, depending on observational constraints at the site such as visibility, weather conditions, and moon phases. When the GRB emission is not visible anymore, observations are typically stopped and further follow-ups are not performed. This study shows that in the case of events similar to GRB\,190114C {and GRB\,221009A}, the best strategy for current or future generation IACTs to probe IGMF would have been to extend the observation of GRBs for at least 3 hours even after the afterglow emission is no longer detected. This can provide the possibility to detect the pair echo emission or, in case of non-detections, to exclude some of the possible magnetic field configuration and provide lower limits on its strength. The range of IGMF strengths that can be studied by current and future generation VHE detectors from GRB follow-ups largely depends on the time delays, the observational conditions and on the intrinsic source properties. This is beyond the scope of this paper.  {A refined study that requires a scan of the parameters involved to give a proper estimate of the IGMF strength is currently under development. Nevertheless, in this first study, we show that a range of IGMF strength values between $10^{-17}$\;G and $10^{-19}$\;G can be explored with GRBs, which makes IGMF studies from (at least) a subclass of GRBs similar to GRB\,190114C competitive with the current limits derived from AGNs.}

Furthermore, we evaluated how the different intrinsic properties of GRBs can modify the detection probability of secondary cascade emission. In particular, we focused on the effect that distance and energetic of GRBs can have on the pair echo emission. The impact of these quantities cannot be easily derived from analytical descriptions since they influence the properties of the mechanisms involved and the behaviour of the photons and pairs. No refined study on these effects has  been performed thus far. In the present work, we considered a GRB with the same intrinsic properties as GRB\,190114C, but with a different redshift (a lower value z $= 0.2$ and a higher z $= 1.0$ value) {and a GRB similar to GRB\,221009A with a higher redshift of $z = 1.0$}. We estimated the resulting pair echo emission and compare it with the MAGIC and CTA-North sensitivities. We prove that despite a nearby event at redshift $z = 0.2$ will have fewer TeV photons reprocessed at GeV energies than the GRB at redshift $z=1.0$, it will generate a much more powerful secondary component. As a result, the power cascade of the reprocessed emission is more sensitive to the intrinsic source properties with respect to the reprocessed emission power.
{Moreover, we prove that in the case of an extraordinary energetic event as GRB\,221009A the CTA-North array would have been able to probe the weak IGMF configuration ($B < 10^{-18}$ G) -- even at a redshift of $z = 1.0,$ which is near the current detection horizon for GRBs at VHE (GRB\,201216C has a redshift of $z = 1.1$).}
Furthermore, this indicates that the current and future generation of extensive air shower detectors, such as HAWC \citep{2017ApJ...843...39A}, LHAASO \citep{2010ChPhC..34..249C}, and SWGO \citep{2022icrc.confE..23H} can play a relevant role in the study of pair echo emission from GRBs. These instruments can indeed cover a wide energy range, starting from $\sim 100$\,s GeV and up to PeV with a sensitivity similar to or better than the CTA-North array above tens of TeV. This will enable studies of both the primary source emission and the secondary delayed pair echo distribution. The sensitivity of these instruments at TeV energies can provide a unique tool for investigating (with unprecedented detail) the shape of the GRB VHE spectrum; in particular, the value of the cut-off that is expected to appear due to the Klein-Nishina suppression and pair-production mechanism around a few tens of TeV (note:\ in this study, we explored two cases, that of 10\,TeV and 50\,TeV). {This was also confirmed by the LHAASO results on GRB\,221009A, where evidence of a cut-off rising at $\sim 13$ TeV was reported}. In addition, extensive air shower detectors can also investigate pair echo emission above hundreds of GeV with a higher duty cycle at the expense of a lower sensitivity with respect to IACTs.

Finally, we show that transient sources (in particular, GRBs) may provide the ideal environment to probe IGMF with gamma-ray emitting sources, overcoming some of the uncertainties that have affected previous studies performed with AGNs. Both objects are indeed expected to be extremely bright, even above 1 TeV. Nevertheless, GRBs, due to their transient nature, fade away on a timescale of a few hours or days in the VHE domain. On the other hand, AGNs are continuous emitters of TeV radiation. This implies that in the case of GRBs, it is possible to identify the duty cycle of these sources; this is a typically unknown value for AGNs but which still happens to strongly affect the results. In addition, the pair-echo emission component in GRBs can be seen at later times with respect to the primary GRB emission and, therefore, it is not affected by the contamination due to the presence of continuous TeV radiation, as in the case of AGNs. This makes GRB IGMF studies more robust and less affected by assumptions on the primary source.

In this study, we did not assume the presence of plasma instabilities during the propagation of the pairs plasma in the intergalactic medium. The electron-positron pairs can indeed experience an additional cooling mechanism before the IC interaction with the CMB, suppressing the cascade emission production. This process was proposed for the first time by \citet{Borderick12} to explain the non-detection of the electromagnetic cascade
in blazar SEDs at GeV energies, as well as the lack of extended emission. Following this first study, there have been numerous subsequent studies aimed at quantifying whether and to what extent plasma instabilities may play a crucial role regarding the development of the cascade in the intergalactic medium (see, e.g., \cite{Miniati14, Schlickeiser13, Sironi14, Chang16, Rafighi17, Vafin18, Vafin19, Batista19, Alawashra22}). Despite many attempts, there has been no commonly accepted view established thus far because the estimation of instabilities strongly depends on the adopted assumptions. However, in case of transient sources such as GRBs or AGN flares, the instabilities might not be a problem after all; indeed, in order to develop, the instabilities require a certain amount of time ($\sim$300 yr, \cite{Borderick12}). In this specific case, the assumed GRB time activity in the VHE bands is the one estimated by MAGIC telescopes; namely, around 40 minutes. The GRB time activity is much lower than the one needed by the instabilities to form \citet{Batista19}, making the IGMF contraints from VHE transients robust.         

\begin{acknowledgements}
{We thank the anonymous referee for the constructive suggestions that helped us to improve the core results of our paper.} D.M. and E.P. acknowledge funding from Italian Ministry of Education, University and Research (MIUR) through the "Dipartimenti di eccellenza” project Science of the Universe. 

D.M. and P.D.V. acknowledge "funding by the European Union - NextGenerationEU” RFF M4C2 project IR0000012 CTA+.
\end{acknowledgements}

\bibliography{bibliography}
\end{document}